\documentclass[fleqn,usenatbib]{mnras}

\usepackage{newtxtext,newtxmath}
\usepackage{subcaption}
\usepackage{soul}
\usepackage[T1]{fontenc}
\usepackage{float}
\DeclareRobustCommand{\VAN}[3]{#2}
\let\VANthebibliography\thebibliography
\def\thebibliography{\DeclareRobustCommand{\VAN}[3]{##3}\VANthebibliography}

\usepackage{graphicx}	
\usepackage{amsmath}	
\usepackage[dvipsnames]{xcolor}

\newcommand{\xv}{{\bf x}}
\newcommand{\kv}{{\bf k}}
\newcommand{\mpch}{h^{-1}\,{\rm Mpc}}
\newcommand{\impch}{h\,{\rm Mpc}^{-1}}

\title[Hitting the Mark]{ Hitting the mark: Optimising Marked Power Spectra for Cosmology}
\author[J. A. Cowell et al.]{
Jessica A. Cowell$^{1,2,3}$\thanks{E-mail: jessica.cowell@physics.ox.ac.uk},
David Alonso$^{1}$, Jia Liu$^{2,3}$\\
$^1$Department of Physics, University of Oxford, Denys Wilkinson Building, Keble Road, Oxford OX1 3RH, United Kingdom\\
$^{2}$Kavli Institute for the Physics and Mathematics of the Universe (Kavli IPMU, WPI), UTIAS, The University of Tokyo, Kashiwa, Chiba 277-8583, Japan\\
$^{3}$Center for Data-Driven Discovery, Kavli IPMU (WPI), UTIAS, The University of Tokyo, Kashiwa, Chiba 277-8583, Japan}

\date{Accepted XXX. Received YYY; in original form ZZZ}

\pubyear{2024}

\begin{document}
\label{firstpage}
\pagerange{\pageref{firstpage}--\pageref{lastpage}}
\maketitle

\begin{abstract}
Marked power spectra provide a computationally efficient way to extract non-Gaussian information from the matter density field using the usual analysis tools developed for the power spectrum without the need for explicit calculation of higher-order correlators. In this work, we explore the optimal form of the mark function used for re-weighting the density field, to maximally constrain cosmology. We show that adding to the mark function or multiplying it by a constant leads to no additional information gain, which significantly reduces our search space for optimal marks. We quantify the information gain of this optimal function and compare it against mark functions previously proposed in the literature. We find that we can gain around $\sim2$ times smaller errors in $\sigma_8$ and $\sim4$ times smaller errors in $\Omega_m$ {compared to using the traditional power spectrum alone}, an improvement of $\sim60\%$ compared to other proposed marks when applied to the same dataset.
\end{abstract}
\section{Introduction}
  Much of the progress made so far in cosmological large-scale structure studies has been using tools that are only optimal for the study of Gaussian random fields, such as correlation functions and power spectra. There are good reasons for these: CMB observations have confirmed that the primordial density fluctuations were very close to Gaussian \citep{Komatsu2003FirstYearWM, 2020_planck}. However, over time, non-linear gravitational clustering leads to highly non-Gaussian features in the small-scale matter distribution. This is the so-called ``cosmic web'', including voids, sheets, filaments, and virialised halos, and its rich non-Gaussian structure may contain invaluable information to constrain different theories of the evolution of the Universe. \citep{astro-ph/0112551,1705.03021}. 

  This non-Gaussian information will not be fully captured by 2-point statistics alone \citep[]{Scoccimarro_1999, Bernardeau_Colombi_Gaztanaga_Scoccimarro_2002}, however there are multiple ways to attempt to recover the extra information, all falling under the umbrella term of ``higher-order statistics'' . One method is to explicitly calculate the higher-order n-point correlations. Most commonly, the next-order correlator to the power spectrum, the bispectrum (3-point correlator), has been investigated and used in measurements of the three-dimensional galaxy distribution \citep{Gil-Marín_Percival_Verde_Brownstein_Chuang_Kitaura_Rodríguez-Torres_Olmstead_2016,2206.02800}. The trispectrum (i.e. 4th-order correlator) has also been exploited \citep{2001Hu,2201.06932,2206.04227}.  The combination of these next-order statistics has been shown to have the power to break degeneracies in cosmological and astrophysical parameters \citep{Bernardeau1997, Pires_Leonard_Starck_2012,Hahn_Villaescusa-Navarro_2021}. The downside to directly calculating the bispectrum or trispectrum is that it can be computationally expensive, and can lead to a large increase in the size of the resulting data vector (with the associated complexity of estimating its covariance matrix and theoretical predictions). However, significant progress has been made at accelerating these calculaions, however \citep{2023Nick,Philcox_Slepian_Hou_Warner_Cahn_Eisenstein_2021, bihalofit, philcox2024polybin3dsuiteoptimalefficient}. There is also no guarantee that this next-order approach leads to an optimal extraction of non-Gaussian information.

  Another approach is to compute summary statistics that capture non-Gaussian information. Compared to n-point correlators, they are often easier to compute and potentially more efficient at capturing non-Gaussian information. An incomplete list includes density split statistics \citep{Gruen_2016, ediss23401, Friedrich_2018}, the integrated cosmic shear 3-point function \citep{Halder_2021, Gong2023CosmologyFT, Halder2023Beyond3C}, counts in cells (CIC) of the PDF \citep{Uhlemann_Friedrich_Villaescusa-Navarro_Banerjee_Codis_2020}, minimum spanning trees \citep{Naidoo_Whiteway_Massara_Gualdi_Lahav_Viel_Gil-Marín_Font-Ribera_2020, Naidoo_Massara_Lahav_2022},  $k$-nearest neighbours \citep{Banerjee_2020}, and wavelet scattering transforms \citep{Cheng2020ANA,Valogiannis2021TowardsAO,Valogiannis_2022, eickenberg2022wavelet, cheng2024cosmologicalconstraintsweaklensing}. There is also significant interest in the advantage of looking at voids as a probe of cosmology \citep{pisani2019cosmic, Davies2020ConstrainingCW}, including the number density of voids \citep{Woodfinden_2022}, and the void size function \citep{Ronconi2017CosmologicalEO,Kreisch2021TheGD,Bayer_Villaescusa-Navarro_Massara_Liu_Spergel_Verde_Wandelt_Viel_Ho_2021, Thiele2023NeutrinoMC}.

  An alternative method is to still use 2-point statistics, but transform the fields themselves such that higher-order information is brought into the two-point correlators. Examples of this include Gaussianised fields \citep{1992Weinberg, Neyrinck_2011, Neyrinck_2011b}, log-normal transformed fields \citep{Neyrinck_2009, Wang:2011fj} and reconstructed density fields \citep{Eisenstein_2007}. In this work, we explore a particular form of these transforms, called the `marked' field, and its associated `marked correlation functions'. Marked correlation functions were first formulated in \cite{1984Stoyan} in the form of marked point processes, where each source was weighted by specific source properties. These were used in measurements of luminosity- and morphology-dependent clustering of galaxies \citep{Beisbart_2000,2021A&A...653A..35S} and its interpretation within the halo model \citep{Sheth_2005,Gottl_ber_2002} and galaxy formation models \citep{Sheth2005Conolly,Skibba_2006}. In the context of cosmological structure formation, marked correlation functions involve locally weighting the original density field by a mark function that depends on the smoothed local density field at the same point. The correlations of this marked field with itself and with the unmarked field then contain information beyond the two-point function of the original field \citep[see e.g.][]{Philcox_2020_pert}. This procedure has several advantages: computing the marked field is inexpensive, and measuring the marked power spectra containing additional non-Gaussian information can be done using the vast existing infrastructure for the estimation of two-point functions, their covariances and, potentially, their theoretical expectations.

  This formalism was more recently popularised by \citet[][W16 hereafter]{White_2016} as a potentially powerful test for modified gravity (MG) theories, and versions of this mark have been used in the literature. The W16 function has been used to test modified gravity theories \citep{Armijo_2018, armijo2023new, Satpathy_2019}, and the effect on MG models in perturbation theory \citep{Aviles_2020}. It has also been compared with other transforms such as the clipping and logarithmic transform \citep{Valogiannis_2018}, as well as the `log mark' and `Gaussian mark' \citep{Hernández-Aguayo_Baugh_Li_2018}. \citet{Massara_2021} demonstrated the power of marked statistics on late-time cosmological parameters using simulations, and galaxy field data \citep{Massara_2023}, finding that they led to a significant improvement in errors on cosmological parameters. Perturbative expressions of the W16 mark within a cosmological context have been derived in \cite{Philcox_2020_pert,2021Philcox}.

  Our aim in this work is to study the possibility of finding an optimal mark function, beyond that allowed by specific functional forms. Specifically, we define ``optimality'' as minimising the final uncertainty on two cosmological parameters, $\Omega_m$ and $\sigma_8$, where $\Omega_m$ is the current non-relativistic energy density fraction, and $\sigma_8$ is the standard deviation of the linear density field on spheres with a radius of $8\,h^{-1}{\rm Mpc}$. Similar studies have been recently carried out in the context finding the optimal mark functions within the context of  modified gravity \citep{kärcher2024optimal}, where they explore various types of mark functions including environmental weightings,  but this paper looks at the we look at the density weighted mark in a cosmological context.

  This paper is organised as follows: in Section \ref{sec:theory}, we describe some of the theories of marked correlation functions and the methodology we use to search for optimal non-parametric mark functions. We present our results in Section \ref{sec:results}, first exploring the potential for optimality of the W16 function, and then finding optimal marks under different conditions, studying their universality. Finally, we summarise our main findings and conclude in Section \ref{sec:conc}.

\section{Theory and Methods}\label{sec:theory}
  \subsection{Marked fields and power spectra}\label{ssec:theory.mark_intro}
    Consider a measurement of an overdensity field $\delta(\xv)$ which, for simplicity, we will assume to be the matter overdensity. Let $\delta_R(\xv)$ be the same field smoothed on a comoving scale $R$. The specific choice of smoothing kernel is arbitrary, and here we will use a Gaussian kernel, with $R$ as its standard deviation. Finally, let $M(\delta_R)$ be the \emph{mark function}, which we will consider to be a smooth, infinitely differentiable function defined over the range of values that $\delta_R$ can take. With these ingredients, the \emph{marked overdensity field} $\Delta(\xv)$ is defined to be
    \begin{equation}
      \Delta(\xv)\equiv M(\delta_R(\xv))\,[1+\delta(\xv)]-\langle M(\delta_R)\,[1+\delta]\rangle,
      \label{eq:mark}
    \end{equation}
    where $\langle\cdots\rangle$ denotes a spatial average over the observed region.
    
    The reasons for considering marked fields are manifold. First, the mark function allows us to up-weight or down-weight different regions based on their local environment (e.g. up-weighting underdensities, where small-scale inhomogeneities may be closer to the linear regime). Moreover, through this simple operation, the marked field becomes a non-linear and mildly non-local function of the overdensity. Thus, the low-order correlations of this field (with itself and with $\delta$), will automatically contain contributions from higher-order correlations \citep{Philcox_2020_pert}. Crucially, we can thus apply well-established techniques used to estimate two-point functions and their uncertainties from real data to extract non-Gaussian information.

    Once $\delta(\xv)$ and $\Delta(\xv)$ have been constructed, we can estimate their three two-point correlators in Fourier space (i.e. power spectra): $P_{\delta\delta}(k)$, $P_{\delta\Delta}(k)$, $P_{\Delta\Delta}(k)$, where
    \begin{equation}
      \langle a(\kv)\,b^*(\kv')\rangle\equiv(2\pi)^3\,\delta^D(\kv-\kv')P_{ab}(k),
    \end{equation}
    and we use the following convention for Fourier transforms:
    \begin{equation}
      a(\kv)\equiv\int\,d^3x\,e^{-i\kv\cdot\xv}\,a(\xv).
    \end{equation}

    The efficiency of a given choice of mark function $M$ should be quantified in terms of its ability to tease out additional information beyond that contained in the power spectrum of the original density field $P_{\delta\delta}(k)$. Given a data vector composed of the three power spectra ${\bf d}\equiv (P_{\delta\delta}(k),P_{\delta\Delta}(k),P_{\Delta\Delta}(k) )$, measured at a set of $k$ values, and assuming that the data follows a multivariate normal distribution (which is a good approximation on sufficiently small scales e.g. \citep{Hamimeche_2008}), we can quantify the amount of information recovered from the Fisher matrix:
    \begin{equation}\label{eq:fisher_matrix}
      \mathcal{F}_{\alpha\beta} = \partial_\alpha{\bf d}^T\,{\sf C}^{-1}\partial_\beta{\bf d},
    \end{equation}
    where $\partial_\alpha\equiv\partial/\partial\theta_\alpha$ denotes the partial derivative of the data vector with respect to the model parameter $\theta_\alpha$, and ${\sf C}$ is the covariance matrix of ${\bf d}$.

    For simplicity, in our analysis, we will only consider two cosmological parameters, the non-relativistic matter fraction $\Omega_m$, and the amplitude of linear matter fluctuations $\sigma_8$. To quantify the information gain, we will use the figure of merit
    \begin{equation}
      {\rm FOM}\equiv {\rm det}(\mathcal{F}).
    \end{equation}
    We will consider both the joint constraints on $\Omega_m$ and $\sigma_8$, in which case $\mathcal{F}$ is a $2\times2$ matrix, as well as either $\sigma_8$ or $\Omega_m$ separately (for which $\mathcal{F}$ is simply a number). We will also compare the FOM for the matter power spectrum $P_{\delta\delta}(k)$ alone, and for the full set of power spectra, in order to quantify the relative information gain when including the marked density field.

  \subsection{Mark functions}\label{ssec:theory.marks}
    The most common form of the mark function used in the literature was proposed by \citet{White_2016}, and takes the form
    \begin{equation}\label{eq:w16_mark}
      M(\delta_R)=\left[1+\frac{\delta_R}{1+\delta_s}\right]^{-p},
    \end{equation}
    where $p$ and $\delta_s$ are free parameters. In this function, $\delta_s$ controls the sensitivity of the mark to the local value of the smoothed overdensity (large positive values of $\delta_s$ make the dependence on $\delta_R$ negligible -- negative $\delta_s$ are not usually considered), and positive/negative values of $p$ upweight under/overdensities.

    In what follows, we will label the function above W16 and, although we will use it as a case study, our main aim is to explore more general functional forms for the mark function, quantifying their optimality.

    \subsubsection{Symmetries of mark functions}\label{sssec:theory.marks.symm}
      \begin{figure}
          \centering
          \includegraphics[width=0.5\textwidth]{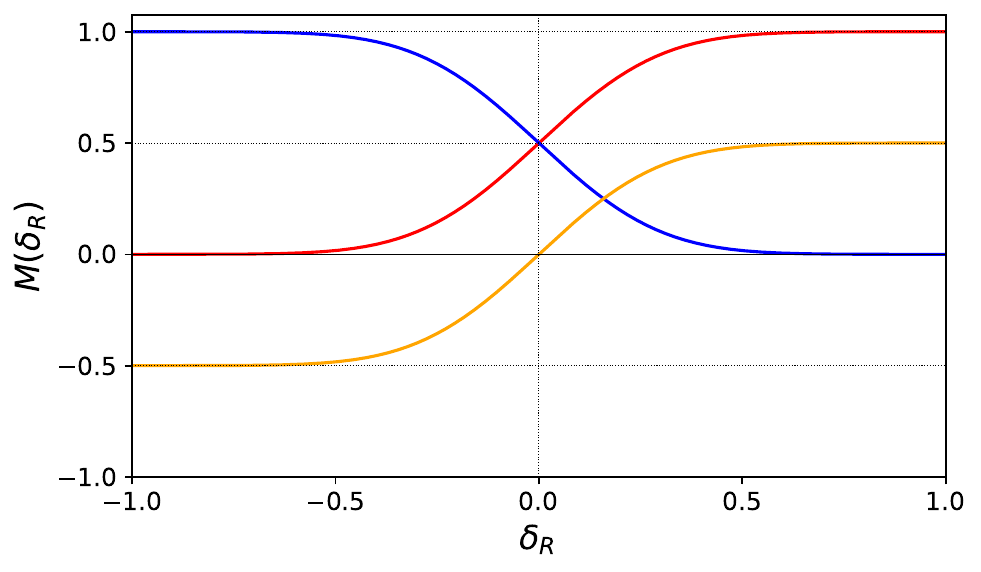}
          \caption{Examples of seemingly distinct (but in fact equivalent) mark functions.}
          \label{fig:marks_symm}
      \end{figure}
      Before we introduce a method to define general mark functions, it is worth exploring its symmetries, in order to avoid spending time exploring functions that are, ultimately, equivalent.

      Two mark functions $M(\delta_R)$ and $M'(\delta_R)$ are equivalent if they are able to recover the exact same amount of information from any given overdensity map (i.e. if they produce the same Fisher information matrix). Under this definition, we can identify two obvious symmetries of mark functions (i.e. transformations connecting two equivalent functions:
      \begin{itemize}
          \item {\bf Linear scaling:} $M'=A\,M$, where $A$ is an arbitrary non-zero constant. The effect of this transformation would be to simply increase the amplitude of the marked overdensity field ($\Delta'=A\,\Delta$) by a known amount, which cannot alter the information encoded in the field. Evidently this symmetry includes also reflections of the mark function across the $x$ axis ($M'=-M$).
          \item {\bf Linear offset:} $M'=M+b$, where $b$ is an arbitrary constant. Under this transformation, the marked field simply receives an additive contribution proportional to the original field: $\Delta'=\Delta+b\delta$. Since our proposed analysis would involve both $\Delta$ and $\delta$, no additional information can be uncovered.
      \end{itemize}
      Mark functions connected via an affine transformation of the form $M'=A\,M+b$ are therefore equivalent. Although the arguments above should suffice, this can be easily proven formally. Under such a transformation, the data vector ${\bf d}$ and its covariance matrix ${\sf C}$ become ${\bf d}'={\sf T}{\bf d}$ and ${\sf C}'={\sf T}{\sf C}{\sf T}^T$\footnote{Note that we are abusing the notation here. The data vector ${\bf d}=(P_{\delta\delta},P_{\delta\Delta},P_{\Delta\Delta})$, is not really a three-dimensional vector, but a concatenation of three power spectra mesured at a set of wavenumbers. Our argument holds nevertheless, by simply promoting the matrix ${\sf T}$ to a block-diagonal matrix.} where the matrix ${\sf T}$ is
      \begin{equation}
        {\sf T}\equiv\left(
        \begin{array}{ccc}
          1 & 0 & 0 \\
          b & A & 0 \\
          b^2 & 2Ab & A^2
        \end{array}
        \right).
      \end{equation}
      The matrix ${\sf T}$ is invertible, and therefore the transformed Fisher matrix is
      \begin{equation}
        {\cal F}'_{\alpha\beta}=\partial_\alpha{\bf d}^T{\sf T}^T({\sf T}^{T})^{-1}{\sf C}^{-1}{\sf T}^{-1}{\sf T}\partial_\beta{\bf d}=\partial_\alpha{\bf d}^T{\sf C}^{-1}\partial_\beta{\bf d}={\cal F}_{\alpha\beta}.
      \end{equation}

      Although this result may seem obvious, it has interesting consequences when trying to interpret a given mark function. Consider, for example, the mark function displayed in blue in Fig. \ref{fig:marks_symm}. Taken at face value, this function would be interpreted as up-weighting underdensities and suppressing overdense regions. Shifting the function downwards by 1, and multiplying the result by -1, we obtain the function shown in red which, instead, favours overdensities and suppresses underdensities. Shifting this by $1/2$ downwards, we instead obtain a function (orange) that upweights both over- and under-densities (the latter with a negative weight), and suppresses regions where the density is close to the mean. Although the interpretation of these three curves as mark functions is clearly different, they are equivalent in terms of the amount of information recovered.

      In order to sift through truly unique mark functions, one may therefore impose a constraint on their amplitude (e.g. by fixing their $L^2$ norm), and by forcing them to cross zero at a specific value of $\delta_R$ (an obvious choice would be $\delta_R=0$, but other values may be used). We will use this to our advantage in the next Section.

    \subsubsection{General mark functions}\label{sssec:theory.marks.gen}
      In order to explore mark functions with a generic form, in search of the optimal one, we make use of Gaussian processes. A Gaussian process is a collection of multivariate normal random variables, representing the values of a function $M(x)$ at different points along the $x$ axis (in our case, $x=\delta_R$). A Gaussian process is thus fully defined by a mean function (which we take to be zero), and a kernel function parametrising the covariance between any two points in the process. In our analysis, we use the radial basis function, also known as ``squared exponential'' kernel. This choice is widely used in machine learning due to its infinite differentiability, which leads to the generation of smooth functions. The kernel is specified by only two parameters: an amplitude $\sigma^2$ and length scale $\ell$:
      \begin{equation}
        k_{\rm SE}(x,x')=\sigma^2\,\exp\left(-\frac{(x-x')^2}{2\ell^2}\right).
      \end{equation}

      With this choice, we thus define a given mark function $M(\delta_R)$ in terms of its values at a set of $N$ fixed points, or nodes, along the $\delta_R$ axis. We label the node positions $\boldsymbol{\delta}_R^*\equiv(\delta^*_{R,1},\cdots,\delta^*_{R,N})$, and the corresponding function values ${\bf M}_*\equiv(M(\delta_{R,1}^*),\cdots,M(\delta_{R,N}^*))$. The value of the function at any other $\delta_R$ is given by the most likely Gaussian process conditioned on the known node values:
      \begin{equation}
        M(\delta_R)={\bf k}^T_*(\delta_R)\,{\sf K}^{-1}_*\,{\bf f}_*,
      \end{equation}
      where the vector ${\bf k}_*(\delta_R)$ and matrix ${\sf K}_*$ are
      \begin{equation}
        k_{*,i}(\delta_R)\equiv k_{\rm SE}(\delta_R,\delta^*_{R,i}),\hspace{12pt}
        K_{*,ij}\equiv k_{\rm SE}(\delta^*_{R,i},\delta^*_{R,j}).
      \end{equation}
      In principle, a general mark function can then be defined in terms of the node values ${\bf M}_*$. We can reduce the number of free parameters, while simultaneously avoiding exploring equivalent functions, taking advantage of the symmetries described in the previous section. To do so, we impose a normalisation constraint on the node values, such that $|{\bf M}_*|^2=1$. In other words, ${\bf M}_*$ is a unit vector that can be described in terms of $N-1$ variables. We choose these variables to be the spherical coordinates parametrising any point in the unit $(N-1)$-sphere. For instance, in the case $N=4$, we would parametrise ${\bf M}_*$ in terms of 3 angles, $(\psi,\theta,\phi)$ as
      \begin{equation}
        {\bf M}_*=(\sin\psi\,\sin\theta\,\sin\phi,
                   \sin\psi\,\sin\theta\,\cos\phi,
                   \sin\psi\,\cos\theta,
                   \cos\psi),
      \end{equation}
      where $\phi\in[0,2\pi)$, and $\{\theta,\psi\}\in[0,\pi)$.

      In our analysis, we will fix the kernel amplitude to $\sigma^2=20$, which is far larger than the typical amplitude of the mark functions given the normalisation imposed above. Finally, the correlation length $\ell$, which governs the smoothness of the resulting function, is chosen based on the inter-node spacing. This itself depends on the dynamic range of $\delta_R$, which varies with the smoothing scale $R$. For the three smoothing scales explored, the corresponding correlation lengths are $\ell(R=5\mpch)=5$, $\ell(R=10\mpch)=2$, and $\ell(R=30\mpch)=0.75$.

\subsection{Finding optimal mark functions}\label{ssec:theory.alg}
  To quantify the performance of different mark functions, we make use of $N$-body simulations. The simulations used are described in Section \ref{sssec:theory.alg.sims}, while the method used to find an optimal mark function from these simulations is outlined in Section \ref{sssec:theory.alg.meth}. Section \ref{sssec:theory.alg.cov} provides further details about the calculation of power spectrum covariances, a central aspect of this analysis.

  \subsubsection{Simulations}\label{sssec:theory.alg.sims}
    We use two types of simulations in this analysis. First, we generate a suite of full $N$-body simulations, each with a box of size $L_{\rm box}=700{\rm Mpc/h}$, with $512^3$ dark matter particles. The simulations are run using the public {\tt Gadget-2} code \citep{Springel:2000yr,Springel_2005}. We run one simulation for a fiducial set of $\Lambda$CDM cosmological parameters
    \begin{equation}
      \{\Omega_m,\sigma_8,h,n_s,\Omega_b\}=\{0.3,0.8,0.7,0.96,0.05\}.
    \end{equation}
    We then run 4 additional simulations, two of them varying $\Omega_m$ by $\Delta\Omega_m=\pm0.02$, while keeping all other parameters fixed, and two of them varying $\sigma_8$ by $\Delta\sigma_8=0.02$. The initial conditions for these simulations were generated using the same seed, in order to partially cancel the effects of cosmic variance when computing the Fisher matrix from their power spectra (as described in Section \ref{sssec:theory.alg.meth}). Snapshots are generated at $z\in\{0,\,0.1,\,0.3,\,0.5,\,1,\,2\}$, although we only study the $z=0$ snapshot here, leaving the investigation of the redshift dependence of the optimal mark to future studies. 

    Furthermore, we generate a suite of 100 fast simulations using the COLA algorithm \citep{Tassev_2013}. These are generated with the same box size and number of particles as the previous simulations, and assuming the fiducial cosmology above. The COLA simulations are used to quantify the impact of non-Gaussian terms in the power spectrum covariance, as described in Section \ref{sssec:theory.alg.cov}.

    All power spectra are estimated from a matter overdensity field calculated using a cloud-in-cell algorithm on a grid with $N_{\rm grid}=256$ cells per side. Power spectra are estimated at a set of wavenumbers covering the range $k\in[0,k_{\rm Ny}]$, with $k_{\rm Ny}=\pi N_{\rm grid}/L_{\rm box}=1.15\,\impch$, in intervals of width $\Delta k=2\pi/L_{\rm box}=0.009\,\impch$.

  \subsubsection{Method overview}\label{sssec:theory.alg.meth}
    In order to quantify the optimality of a given mark function, defined in terms of the Gaussian process nodes (or rather, the associated hypersphere angles defined in Section \ref{sssec:theory.marks.gen}), we do the following:
    \begin{itemize}
      \item Having measured the original overdensity field $\delta$, and its smoothed version $\delta_R$, we evaluate the mark function on the latter, and construct the marked overdensity as defined in Eq. \ref{eq:mark}. This is done for all 5 simulations (fiducial cosmology and 1-parameter departures).
      \item We then compute the three auto- and cross-power spectra from these fields and build the complete data vector ${\bf d}\equiv\{P_{\delta\delta}(k),\,P_{\delta\Delta}(k),\,P_{\Delta\Delta}(k)\}$ for each simulation.
      \item We construct the data vector derivatives via finite differences as
      \begin{equation}
        \partial_\alpha{\bf d}\simeq\frac{{\bf d}(\theta_\alpha+\delta\theta_\alpha)-{\bf d}(\theta_\alpha-\delta\theta_\alpha)}{2\delta\theta_\alpha},
      \end{equation}
      where ${\bf d}(\theta)$ is the vector of power spectra estimated from the simulation with parameters $\theta$.
      \item We construct an approximate covariance matrix, in the form of the analytical disconnected (or ``Gaussian'') covariance, estimated from the power spectra of $\delta$ and $\Delta$ in the fiducial simulation. See Section \ref{sssec:theory.alg.cov} for a detailed discussion of the covariance matrix.
      \item We combine the results of the last two points to compute the Fisher matrix (Eq. \ref{eq:fisher_matrix}), and the figure of merit associated with this mark.
    \end{itemize}

    The steps above allow us to estimate the figure of merit for a given mark function, defined by the hypersphere angles that determined its Gaussian process nodes. To find the optimal mark function, we thus maximise this figure of merit as a function of the node angles. In our fiducial analysis, we use $N=4$ nodes, and therefore we marginalise over 3 angle parameters. The location of the nodes in the $\delta_R$ axis depends on the value of $R$ used (which governs the dynamic range of $\delta_R$). We spread the 4 nodes evenly between the minimum and maximum  $\delta_R$ values. We show the information for the 3 smoothing cases in Table \ref{tab:nodes_info}, and we will study the dependence of our results on these choices in Section \ref{sssec:stab}. We use the modified Powell algorithm as implemented in {\tt scipy} \citep{2020SciPy-NMeth} to find the optimal mark function. In practice, we do this by minimising the negative logarithm of the FOM, which is a smoother function of the node parameters. In order to avoid local minima, and to study the possibility of finding distinct, but equally performing mark functions, we repeat this minimisation starting from a set of 100  random points on the hypersphere and study the resulting distribution of optimal mark functions.

    \begin{table}
        \centering
        \begin{tabular}{c|c|c}
             R ( $h^{-1}$MPc)& $\delta_R(min, max)$  & $\ell$ \\
             \hline
                  
              5 &-0.9, 18.8 & 5.0\\
              10 &-0.8, 4.6 & 2.0\\
              30 & -0.4,0.5& 0.75\\
        \end{tabular}
        \caption{Node spacing information for calculating mark functions. R corresponds to the smoothing scale, while we also show the range of $\delta_R$ between which our 4 nodes are placed, and $\ell$ corresponds to the length scale of the Gaussian process.}
        \label{tab:nodes_info}
    \end{table}

  \subsubsection{Covariance matrices}\label{sssec:theory.alg.cov}
    A crucial aspect of our analysis is the power spectrum covariance matrix ${\sf C}$ used to calculate the Fisher matrix and figure of merit. Since the covariance of $P_{\delta\Delta}(k)$ and $P_{\Delta\Delta}(k)$ depends on the choice of mark function, this matrix must be recalculated for every value of the node parameters explored in the minimisation procedure described in the previous section. Calculating this covariance accurately from simulations at every step would be prohibitively expensive, and therefore we employ an approximate estimate covariance matrix. Specifically, we calculate the disconnected component of the covariance matrix, equivalent to the covariance matrix assuming that all fields involved ($\delta$ and $\Delta$) are Gaussian. In this case, the covariance between the power spectra of a pair of fields $(a,b)$ and another pair $(c,d)$ is
    \begin{equation}
      {\rm Cov}(P_{ab}(k),P_{cd}(k'))=\frac{\delta^K_{k,k'}}{N_k}\left[P_{ac}(k)P_{bd}(k)+P_{ad}(k)P_{bc}(k)\right],
    \label{eq:gauss_cov}
    \end{equation}
    where $\delta^K_{kk'}$ is the Kronecker delta, and $N_k$ is the number of modes used to calculate the power spectra in the bin of wavenumbers labelled by $k$. The disconnected covariance has been found to be the dominant contribution to the matter power spectrum covariance, especially on mildly non-linear scales \citep{1607.00043}. Therefore, although not accurate in detail, it should provide a reasonable estimate of the power spectrum uncertainties when comparing the performance of different mark functions. Since the validity of this approximation is less well studied for general marked fields and, in general, to mitigate the impact of its inaccuracies on our results, we limit our analysis to scales $k<0.3\impch$ and will study the impact of this choice on our results.

    Once the optimal mark function has been determined using this approximation, we make use of the COLA simulations to compute a better estimate of the covariance matrix,
    \begin{equation}
     {\rm Cov}_{ij}= {\rm Cov}^{\rm NG, COLA}_{ij} \sqrt{\frac{{\rm Cov}^{\rm G,N-body}_{ii}{\rm Cov}^{\rm G,N-body}_{jj}}{{\rm Cov}^{\rm G,COLA}_{ii}{\rm Cov}^{\rm G,COLA}_{jj}}},
     \label{eq:final_cov}
    \end{equation}
    where ${\rm Cov}^{\rm G}$  is the Gaussian covariance estimated from the measured power spectra (either in the COLA or true N-body sims), following Eq. \ref{eq:gauss_cov}, and ${\rm Cov}^{\rm NG,COLA}$ is the full covariance matrix calculated from the COLA realisations. The motivation for this equation stems from the fact that we expect the COLA realisations to lack power on small scales when compared to the true N-body simulation. Thus we scale the non-Gaussian COLA covariance by the relative amplitude between the Gaussian covariance estimated from both types of simulations, along the diagonal. Once calculated, we use this covariance to recalculate the FOM associated with the optimal mark function and quantify the relative information gain compared with that predicted using the analytical disconnected covariance.

    It is important to note that, for a general mark function, the fields $\delta$ and $\Delta$ (and therefore their power spectra) may be highly correlated. This is obvious, for example, in the case of a constant mark $M(\delta_R)={\rm const.}$, for which the fields are completely correlated, and no new information can be gained from $\Delta$. In these cases, the combined power spectrum covariance ${\sf C}$ may be very close to singular, and hence numerically unstable upon inversion (e.g. in the constant mark case, two-thirds of the covariance matrix eigenvalues are exactly zero, since $P_{\Delta\Delta}\propto P_{\Delta\delta}\propto P_{\delta\delta}$). To avoid this issue, we construct a pseudo-inverse of our covariance matrix using its singular value decomposition. We calculate the eigenvalues and eigenvectors of the matrix, identify all eigenvalues lower than a fraction $f_{\rm thr}$ of the largest eigenvalue, and construct the inverse matrix by inverting all eigenvalues except those below the threshold, which we set to zero in the pseudo-inverse. This effectively removes all information from modes of the data vector that are aligned with the affected eigenvectors. As in \citet{Park_2023}, we use a condition number, of $f_{\rm thr}=10^{-7}$ in our analysis.
     
\section{Results}\label{sec:results}

\subsection{Exploring the W16 mark}\label{ssec:w16_results}

  \begin{figure}
    \centering
    \includegraphics[width=\columnwidth]{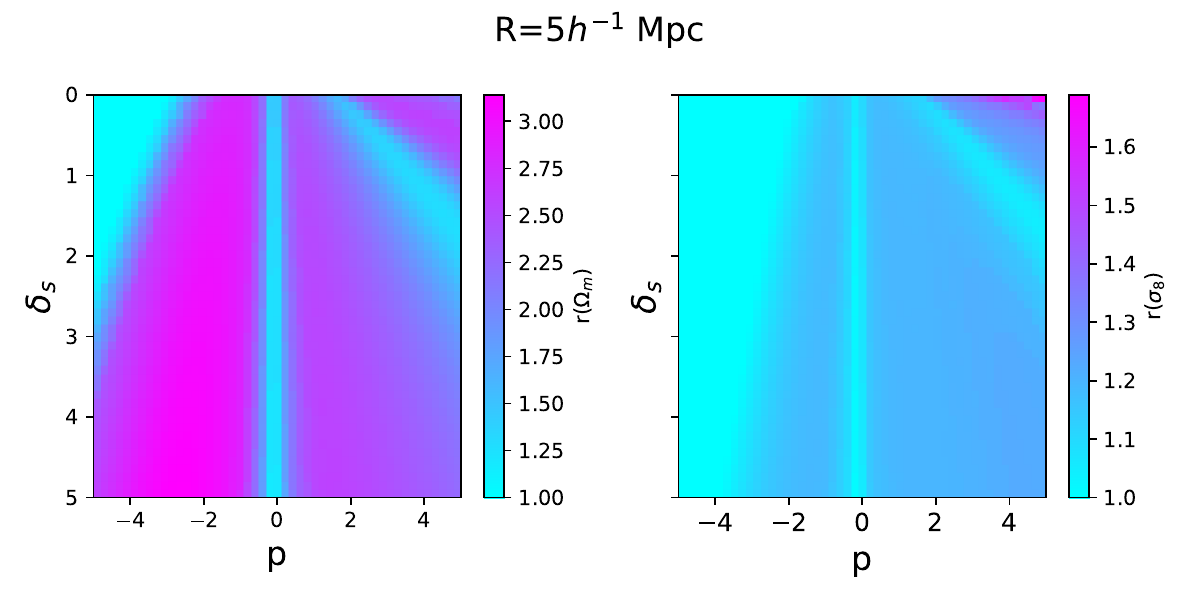}
    \includegraphics[width=\columnwidth]{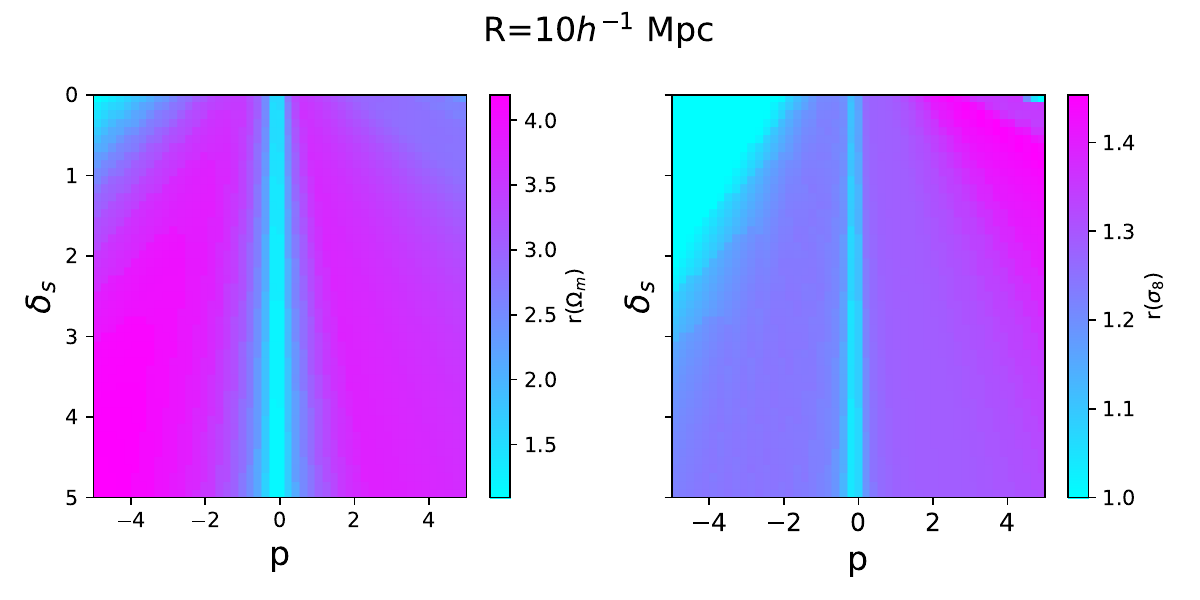}
    \includegraphics[width=\columnwidth]{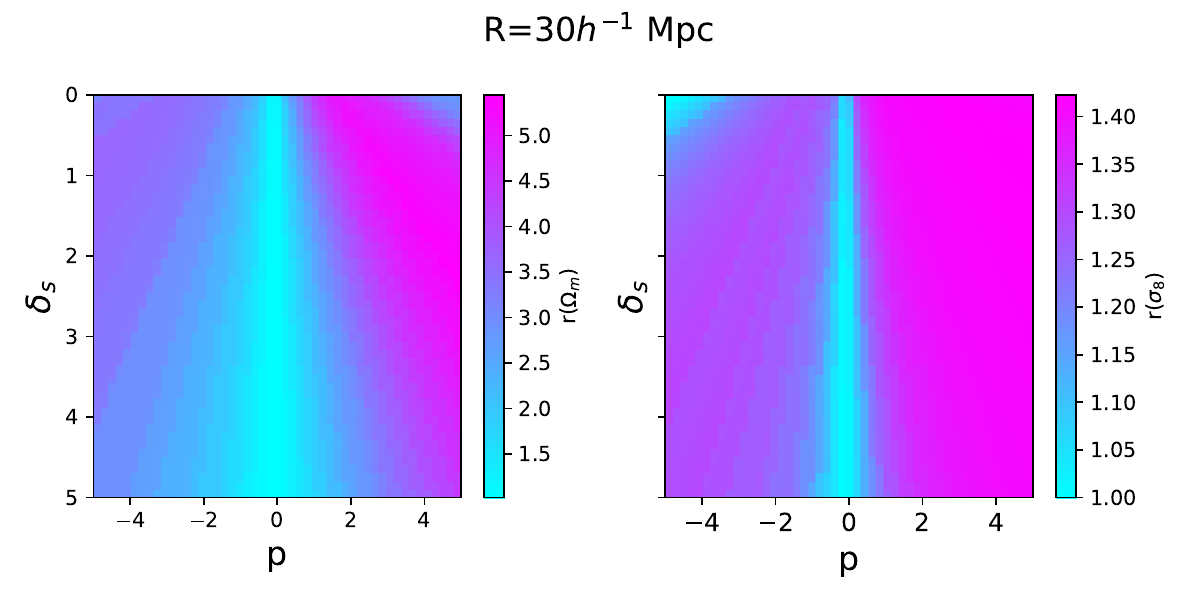}
    \caption{Error improvement (see Eq. \ref{eq:rimp}) in $\Omega_m$ (left) and $\sigma_8$ (right) when altering the free parameters $\delta_s$ and $p$ of the W16 mark function in Eq. \ref{eq:w16_mark}. We also calculate for 3 different values of the smoothing scale, $R$, of (5, 10, 30) $h^{-1}$Mpc.}\label{fig:w16_fish_plots}
  \end{figure}

  The performance of the W16 mark function for different values of its free parameters $(\delta_s,p)$, and for different smoothing scales was studied in \citet{Massara_2021,Massara_2023,2024SimbigMarks}. They find the values to depend on the type of data used, N-body simulations, and whether galaxy or matter overdensity fields were used. In \citet{Massara_2021}, out of 125 combinations of $R$, $\delta_s$, and $p$ , the best parameters were found to be $R = 10 h^{-1}$ Mpc, $p = 2$, and $\delta_s = 0.25$ using N-body simulations. In \cite{Massara_2023} $60$ parameterisations were explored, finding slightly different marks with $R\simeq30 h^{-1}$Mpc, $p=1$ and $\delta_s$ between 0.1 and 0.5. The cases $(R,p,\delta_s)=(30\,\mpch,0.5,1.0)$ and $(R,p,\delta_s)=(15\,\mpch,0.5,0.1)$ were employed on real data in \cite{2024SimbigMarks} using simulation-based inference. We later compare our results to marks using these parameters. 

  The fast, simplified evaluation procedure described in the previous section allows us here to quantify the performance of the W16 mark function on a finer parameter grid. Specifically, for 3 different Gaussian smoothing scales of $R=5,\,10$, and $30\,\mpch$\footnote{Note that these are not directly comparable to the smoothing scales used in \citet{Massara_2021, Massara_2023} as they used top-hat window functions.}, we vary $p$ between (-5, 5), and $\delta_s$ between (0,5), on a $49\times49$ grid (i.e. $2{,}401$ evaluations for each smoothing scale). We then calculate the parameter errors from the Fisher matrix (see Eq. \ref{eq:fisher_matrix}) $\sigma(\theta_i)=\sqrt{{{\mathcal{ F}}^{-1}}_{ii}}$, where $\theta_i\in\{\Omega_m,\sigma_8\}$. We calculate this error in two cases: for a data vector including the three power spectra $(P_{\delta\delta}(k),P_{\delta\Delta}(k),P_{\Delta\Delta}(k))$, and including only $P_{\delta\delta}(k)$. Finally, we estimate the error improvement factor $r$ as the ratio of both errors:
  \begin{equation}\label{eq:rimp}
    r(\theta)=\frac{\sigma(\theta|\delta\delta)}{\sigma(\theta|\delta\delta+\delta\Delta+\Delta\Delta)}.
  \end{equation}

  The results of this exercise are shown in Figure \ref{fig:w16_fish_plots}, where the smoothing scale increases from top to bottom, and error improvements for $\sigma_8$ and $\Omega_m$ are plotted on the left and right columns, respectively. We see the highest improvement for $\Omega_m$ at a smoothing radius of $30\,\mpch$, contrasting to $\sigma_8$ at $R=5\,\mpch$. We can also observe large degeneracies in the mark function parameters for a given smoothing scale (i.e. broad regions of the $(p,\delta_s)$ space achieving roughly the same error improvement). For example, in the case of $\Omega_m$, a broad region with $p\lesssim-1$, largely independent of $\delta_s$ can achieve close-to-optimal constraints for $R=5$ and $10\,\mpch$ (interestingly, this trend changes sharply to positive values of $p$ for $R=30\,\mpch$). In contrast, positive values of $p$ seem to achieve close to optimal constraints on $\sigma_8$, almost independently of $\delta_s$. In all cases, the error improvement ratio drops sharply near $p=0$, where the mark function is approximately constant, as expected. \citet{Massara_2023} found maximum information recovery for at $p=1$ at $R=30\,\mpch$, and $p=2$ for $R=10\,\mpch$, using a top-hat smoothing radius. Qualitatively, we can see similarities, however, there is ultimately no clear ``best'' mark, or clear local maxima and minima in the $(p,\delta_s)$ plane. This motivates our goal to explore more general mark functions.
  \begin{figure*}
    \includegraphics[width=\textwidth]{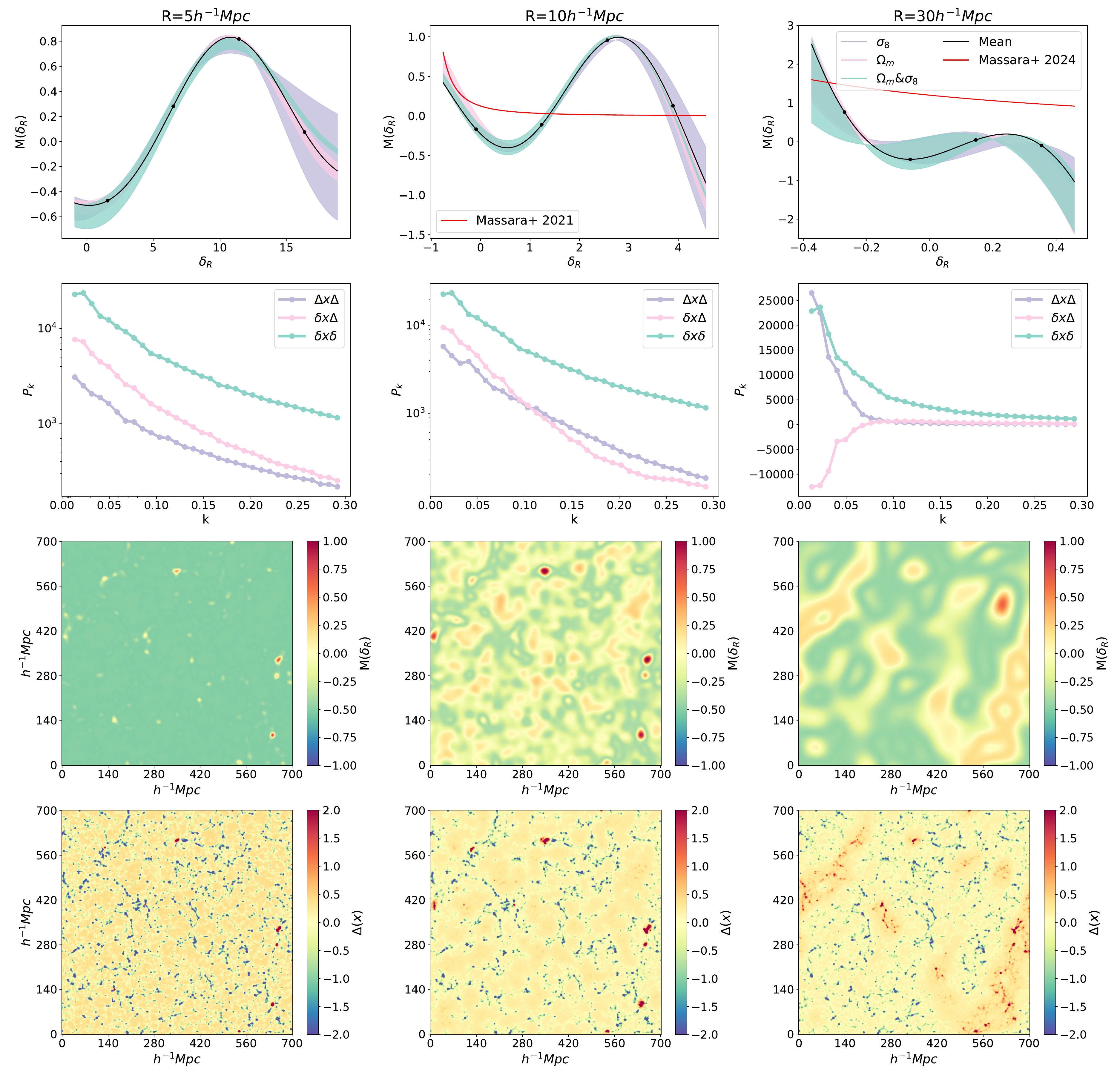}
    \caption{The columns correspond to results for different smoothing scales, corresponding to $R =$ 5, 10, 30 $h^{-1}$Mpc respectively. In the first row, we display the optimal mark functions $M(\delta_R)$ found for 900 optimisers, 300 for each column, as a function of the value of the smoothed over-density. The marks are colour-coded for each of 3 different figures of merits (FOM), with the filled region illustrating the range of different curves found by the optimisers (i.e. maximum and minimum). 
    The second row shows the power spectra for the autocorrelation and cross-correlation of the density field, $\delta$, and marked density field, $\Delta$, for the mark functions shown above. The third row shows a slice of the mark function as a function of the density field of the simulation $M(\delta_R)$, highlighting the scales of features each mark upweights/downweights. The marked field $\Delta({\bf x})$ is shown in the final row, which is simply a product of the marked field in row 3 and the fiducial field in Figure \ref{fig:og_field}.}
    \label{fig:optimisers}
  \end{figure*}

  We find a maximum error improvement ratio of $\sim1.4$ in $\sigma_8$ and $\sim 5$ in $\Omega_m$. Note, however, that these error improvement ratios were estimated using the Gaussian covariance approximation from Eq. \ref{eq:gauss_cov}, and should therefore not be taken at face value. As shown in Table \ref{tab:optimisers}, the effect of this approximation can be significant, with the error improvement ratios changing to $1.7$ and $2.8$, respectively. Nevertheless, this exercise demonstrates the difficulty in finding an optimal mark from a family of functions with limited freedom.

\subsection{Optimising the mark function}\label{ssec:optimal_results}

  \begin{table}
    \begin{center}
    \begin{tabular}{ |c|c|c|c||}
      Mark type & $R\,[\mpch]$& ${r}(\sigma_8)$  & ${r}(\Omega_m)$\\ 
      \hline
      \hline
                               &  5 & 1.3 & 2.7 \\ 
      Optimised, Gaussian Cov. & 10 & 1.4 & 4.1 \\ 
                               & 30 & 1.4 & 4.5 \\ 
      \hline
                               &  5 & 2.0 & 4.5 \\ 
      Optimised, Combined Cov. & 10 & 2.1 & 4.6 \\ 
                               & 30 & 2.0 & 3.5 \\ 
      \hline
      W16, Gaussian Cov.       & 30 & 1.4 & 4.5\\ 
      W16, Combined Cov.      & 30 & 1.7 & 2.8\\
      \hline
    \end{tabular}
    \end{center}
    \caption{Results of error improvement from the 900 optimised mark functions shown in Figure \ref{fig:optimisers}. We display the error ratio, $r = \epsilon^\Delta_{i}/{\epsilon_{i}}$ of the mean curves displayed in Figure \ref{fig:optimisers}, with the various covariance approximations described in Section \ref{sssec:theory.alg.cov}. The last two rows show the results obtained with the W16 parametrisations for parameters $p=1$, $\delta_s=0.5$ \citep{2024SimbigMarks}.}\label{tab:optimisers}
  \end{table}
    \begin{figure}
        \centering
        \includegraphics[width=0.5\textwidth]{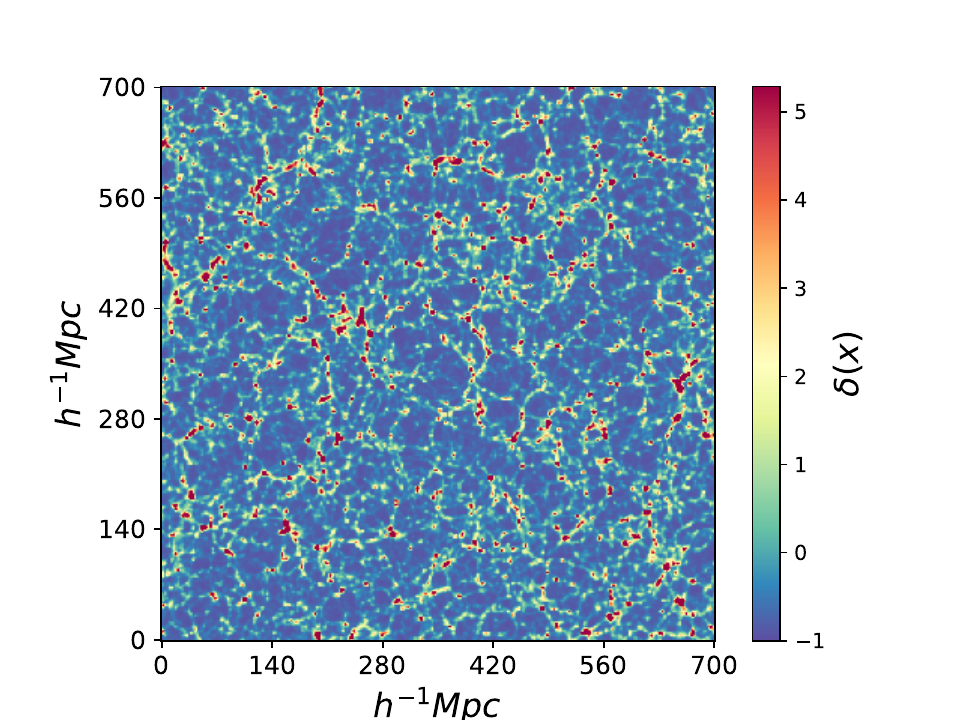}
        \caption{A slice of the simulation for the original, not marked, over-density of the fiducial cosmology. Values are capped at $2 \sigma$ values away from the mean value for clarity, where $\sigma$ is the standard deviation of the field.}
        \label{fig:og_field}
    \end{figure}
  \begin{figure*}
    \centering
    \includegraphics[width=\textwidth]{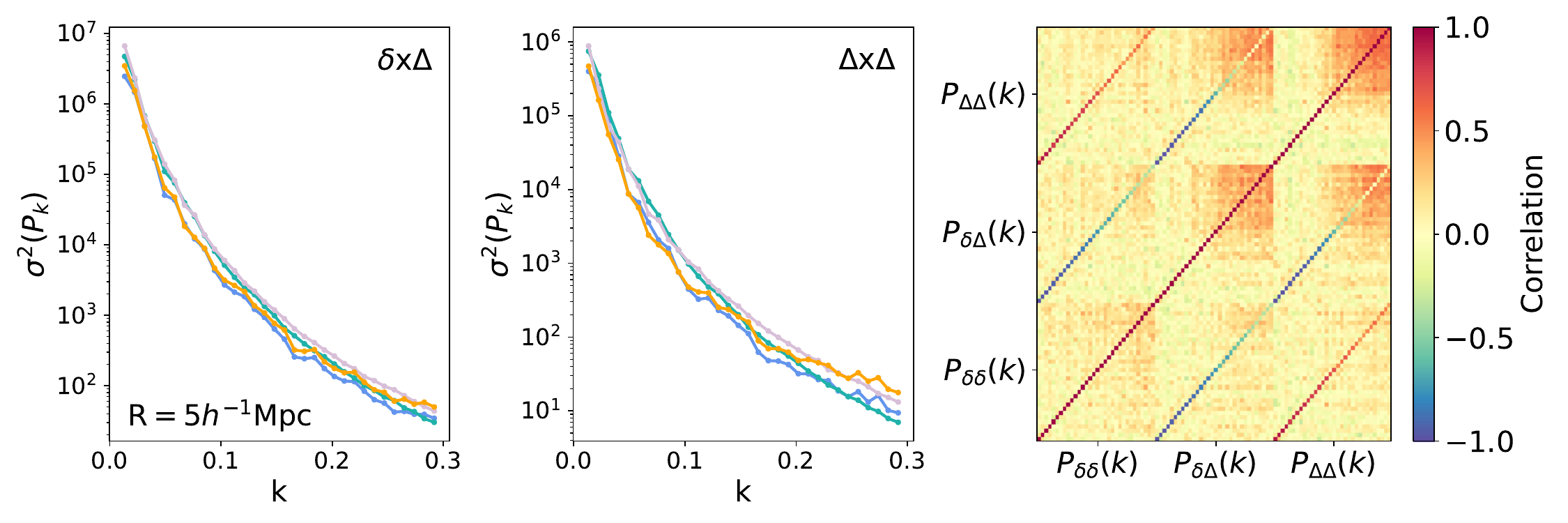}
    \includegraphics[width=\textwidth]{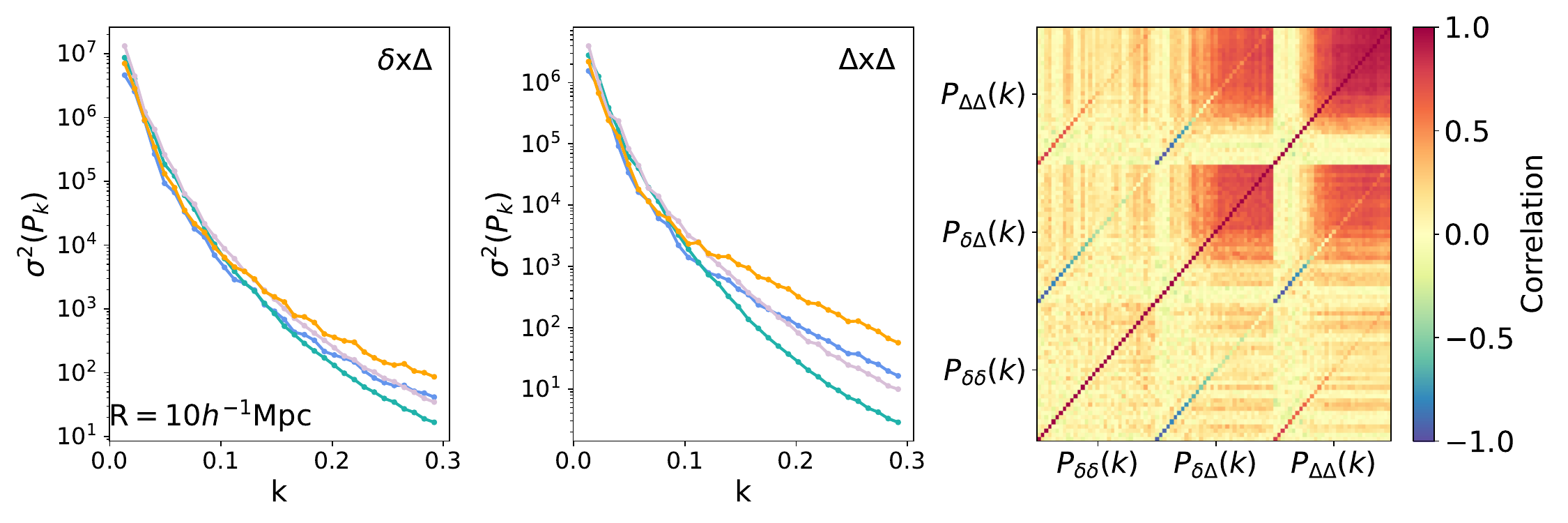}
    \includegraphics[width=\textwidth]{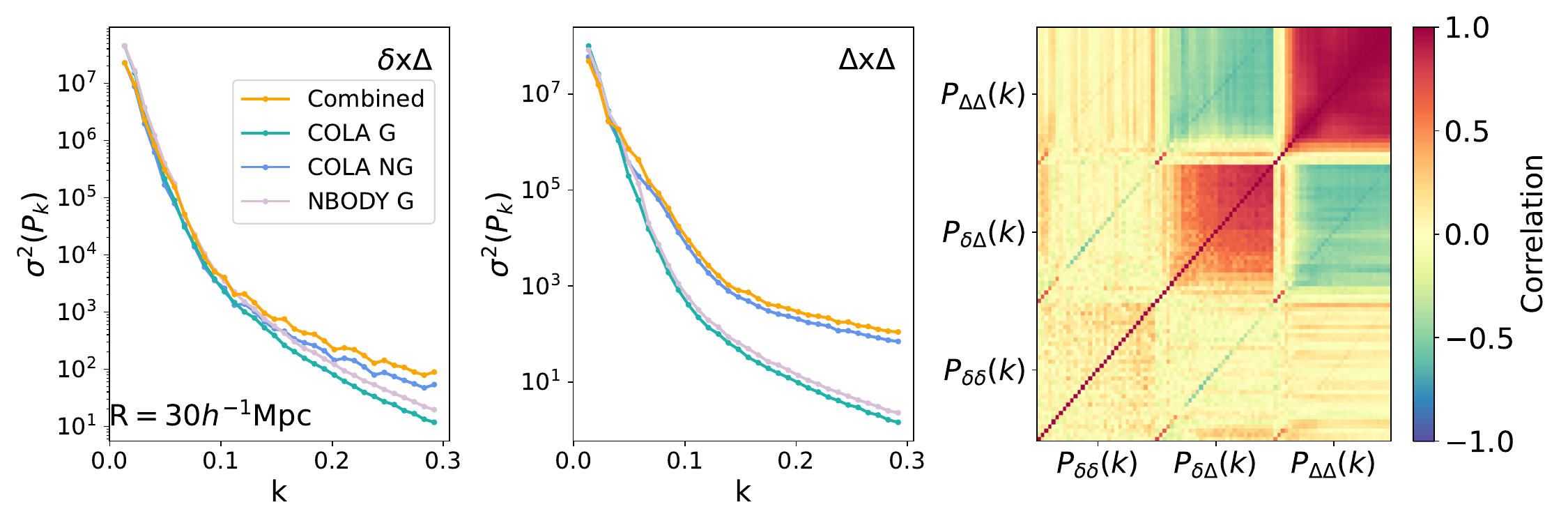}
    \caption{Diagonals of the various covariance matrices for the 3 mark functions for smoothing scales R=5$h^{-1}$Mpc, 10$h^{-1}$Mpc, and 30$h^{-1}$Mpc from top to bottom respectively. We plot the Non-Gaussian (NG) Covariance calculated from the 100 COLA simulations, compared to the Gaussian approximation from Eq. \ref{eq:gauss_cov} calculated for the full N-body simulations, and our combined covariance estimate from Eq. \ref{eq:final_cov} In the 3rd column we show the corresponding correlation matrices.}\label{fig:cov_plots}
  \end{figure*}
  \subsubsection{Optimal marks}\label{sssec:optimal}
    Given the degeneracies displayed by the W16 mark function, it is important to ensure that, for the more complex Gaussian process marks we aim to optimise, we can find true, converged, global minima. To do so, for a given choice of smoothing scale and figure of merit, we run 100 optimisations starting from random points on the surface of the 4D sphere. Once the minimisers have converged, we can flag those that ended up in local minima, having achieved a significantly lower figure of merit than the optimum. For the 3 smoothing scales and FOMs explored (i.e. optimising for $\Omega_m$ or $\sigma_8$ separately, and for their combined FOM), this leads to a total of 900 mark functions.

    The results of this study are summarised in Figure \ref{fig:optimisers}. Each column corresponds to a different smoothing scale, $R = 5,\,10$, and  $30\,\mpch$  from left to right. The first row shows the optimised mark function for each case, each containing the 300 mark curves found by the optimiser, 100 for each FOM. Each FOM is plotted in a different colour, with the filled region illustrating the range of different curves found by the optimisers (i.e. maximum and minimum). Remarkably, for each smoothing scale, we find that the minimiser converges on almost the same mark function for all three FOM choices. Consequently, it seems possible to find a single function that optimises constraints on both of these parameters simultaneously. The black line shows the mean mark function, calculated from the 900 optimisers. The physical interpretation of this optimal mark is not immediately evident, especially given the additive and multiplicative symmetries discussed in Section \ref{sssec:theory.marks.symm}, which would allow us to shift the mark functions shown in the figure up or down by an arbitrary amount. Since these symmetries allow us to fix the value of the mark function at a given $\delta_R$, let us take $M(\delta_R=0)=0$ as a natural choice (this ensures that, to lowest order, the marked field is at least quadratic in the matter overdensity). With this constrains in mind, consider for concreteness the optimal mark for $R=10\,\mpch$: since the mark function reaches a minimum close to $\delta_R=0$, it effectively upweights both underdensities and overdensities away from the mean \emph{with the same sign}\footnote{Note that, under the multiplicative symmetry, both positive and negative mark function values can be interpreted as upweighting the overdensity field in the corresponding regions, although the relative sign between different parts of the mark function can be used to unlock additional information.}. The other smoothing scales seem to approximately reproduce this structure (a minimum close to $\delta_R=0$). This is a markedly different behaviour to that of the W16 function, shown as a red line in the Figure for the parameters taken from \citep{Massara_2021, 2024SimbigMarks} of $R=10, p = 2,
    \delta_s = 0.25 $ and $R=30, p = 1,
    \delta_s = 0.5$ respectively. Since the W16 function is monotonically decreasing, even if we used the shift symmetry to enforce the constraint $M(\delta_R=0)=0$, the point $\delta_R=0$ would simply be a zero crossing and not a minimum. With our normalisation, the W16 function would thus weigh under- and over-densities with opposite signs.
  
    The second row in Fig. \ref{fig:optimisers} shows the three power spectra $\{P_{\delta\delta}(k),P_{\delta\Delta}(k),P_{\Delta\Delta}(k)\}$ for the optimal mark function in each case. It is interesting to note that while, with our choice of sign for the mark function, all three power spectra are positive for $R=5$ and $10\,\mpch$, the cross-correlation $P_{\delta\Delta}(k)$ changes sign as a function of scale for $R=30\,\mpch$. This is not necessarily surprising, since the morphology of the regions over which the smoothed overdensity field takes different values (and hence the structures that get positively or negatively weighted) is a strong function of the smoothing scale. It is worth noting that, although the power spectra used for this analysis were measured from a single N-body simulation, they are numerically stable, as are the finite-difference partial derivatives used to compute the Fisher matrix. This is greatly aided by the fact that the N-body simulations were run with different cosmologies all using initial conditions generated from the same seed. Therefore we expect the qualitative results found here to be robust against numerical instabilities in the method used to estimate the Fisher matrix and associated parameter uncertainties. 

    This can be seen in the third and fourth rows of the same figure, which show the mark function evaluated on the smoothed overdensity field, $M(\delta_R({\bf x}))$, and the marked field $\Delta({\bf x})$, defined in Eq. \ref{eq:mark}, in a 2D slice of the N-body simulation used here. The original overdensity field is shown in Fig.\ref{fig:og_field}. The marked fields display, in all cases, a significantly different morphology, with some structures in the original field highlighted and others suppressed. The marked fields, in which extreme environments (e.g. large overdensities or voids) can undergo significant enhancements, can therefore display very significant non-Gaussian fluctuations, potentially stronger than those of the original unsmoothed overdensity. This may play an important role in the covariance matrix of the resulting power spectra, as we will see.

    The error improvement factors resulting from these optimal marks, calculated as described in Section \ref{ssec:w16_results}, are shown in Table \ref{tab:optimisers}. The table displays results for two different choices of the covariance matrix: the Gaussian approximation of Eq. \ref{eq:gauss_cov} (first three rows for each smoothing scale), used to obtain the optimal mark function, and the combined covariance of Eq. \ref{eq:final_cov}, calculated from the COLA simulations and accounting for non-Gaussian contributions (rows 4-6). Rows 7 and 8 show the improvement factors found for both covariances using the W16 function with close-to-optimal parameters $p=1$, $\delta_s=0.5$ \cite{2024SimbigMarks}. We see that the choice of covariance has a relevant impact on the final achieved constraints, with $r(\sigma_8)$ changing by up to 50\%, and $r(\Omega_m)$ by up to 60\%. The largest improvement factor is achieved for the $10\,\mpch$ smoothing scale, which achieves an error reduction factor of 2.1 and 4.6 for $\sigma_8$ and $\Omega_m$, respectively (compared to 1.4 and 4.1 for the Gaussian covariance). We find that this optimal mark function outperforms the W16 function, by $\sim 20\%$ for $\sigma_8$, and by  $\sim 60\%$ for $\Omega_m$. Fixing the smoothing scale to the same one used for the W16 function ($30\,\mpch$), these drop to $\sim 20\%$ and $\sim 25\%$, respectively. The results found thus far therefore confirm that it is possible to find more general marks that outperform simple parametrisations such as W16, and that the resulting mark is relatively universal, in that it can optimise the constraints on more than one parameter.

  \subsubsection{Covariance matrices}\label{sssec:covs}
    Given the non-negligible impact precision with which the covariance matrix is estimated on the final constraints, it is worth exploring the potential origin of the observed differences in more detail. These are displayed in Fig. \ref{fig:cov_plots}. The three rows show results for the $R=5,\,10,$ and $30\,\mpch$ smoothing scales. In each row, the first and second plots show the diagonals of the covariance matrix for the $P_{\delta\Delta}(k)$ and $P_{\Delta\Delta}(k)$ power spectra, respectively. Results are shown for the Gaussian covariances estimated from the true N-body simulations and from COLA, for the covariance estimated directly from the COLA realisations, and for the combination used in our analysis (see Eq. \ref{eq:final_cov}). The third column then shows the correlation matrix associated with the COLA covariance, with the blocks corresponding to different pairings of the measured power spectra $\{P_{\delta\delta},P_{\delta\Delta},P_{\Delta,\Delta}\}$ clearly visible and marked.
  
    First of all, comparing the Gaussian errors in the first two columns, we can confirm our expectation that the COLA realisations recover a reduced clustering amplitude on small scales, which motivates the correction included in Eq. \ref{eq:final_cov}. Secondly, we can see that the full covariance estimated from COLA displays a significantly larger variance on small scales than that predicted by the Gaussian approximation, particularly for the $R=10$ and $30\,\mpch$. As shown in the third column of the figure, this large non-Gaussian contribution is highly correlated across different wavenumbers, dominating the off-diagonal elements of the covariance on small scales. Effectively, this seemingly large contribution is equivalent to a large increase in the statistical uncertainties of a single mode (or at most a small number of them), which explains its relatively moderate impact in the final constraints ($40\%$ at most) in spite of its large contribution to the small-scale power spectrum uncertainties.

    The presence of this correlated contribution, and its growing amplitude for larger smoothing scales, is to be expected. The marked field is effectively a product of two fields: $m({\bf x})\equiv M(\delta_R({\bf x}))$, and $\delta({\bf x})$ where, by construction, $m({\bf x})$ only has power on scales $k\,\ll 1/R$. Therefore, on small scales ($kR\gg1$):
    \begin{itemize}
      \item $m$ acts effectively as a constant (but stochastic) amplitude ($\Delta({\bf x})\sim m\,\delta({\bf x})$).
      \item $m$ and $\delta$ are largely statistically uncorrelated.
    \end{itemize}
    In this regime, the power spectrum of $\Delta$ can therefore be approximated as
    \begin{equation}
      \hat{P}_{\Delta\Delta}(k)=m^2\,\hat{P}_{\delta\delta}(k),
    \end{equation}
    and therefore, taking into account the stochasticity of $m$, its covariance is:
    \begin{align}
      &{\rm Cov}(\hat{P}_{\Delta\Delta}(k), \hat{P}_{\Delta\Delta}(k'))\\
      &\hspace{22pt}\equiv\langle\hat{P}_{\Delta\Delta}(k)\hat{P}_{\Delta\Delta}(k')\rangle-\langle\hat{P}_{\Delta\Delta}(k)\rangle\langle\hat{P}_{\Delta\Delta}(k')\rangle\\
      &\hspace{22pt}\simeq\langle m^4\rangle{\rm Cov}(\hat{P}_{\delta\delta}(k),\hat{P}_{\delta\delta}(k'))+\sigma_{m^2}^2P_{\delta\delta}(k)P_{\delta\delta}(k'),
    \end{align}
    where $\sigma_{m^2}^2\equiv\langle m^4\rangle-\langle m^2\rangle^2$. The second term in the equation above, due to the stochasticity of the marked smooth field, is 100\% correlated on all scales and effectively allows additional variance for the power spectrum to vary its amplitude while preserving its shape.

  \subsubsection{Stability to analysis choices}\label{sssec:stab}
    \begin{figure}
      \centering
      \includegraphics[width=\columnwidth]{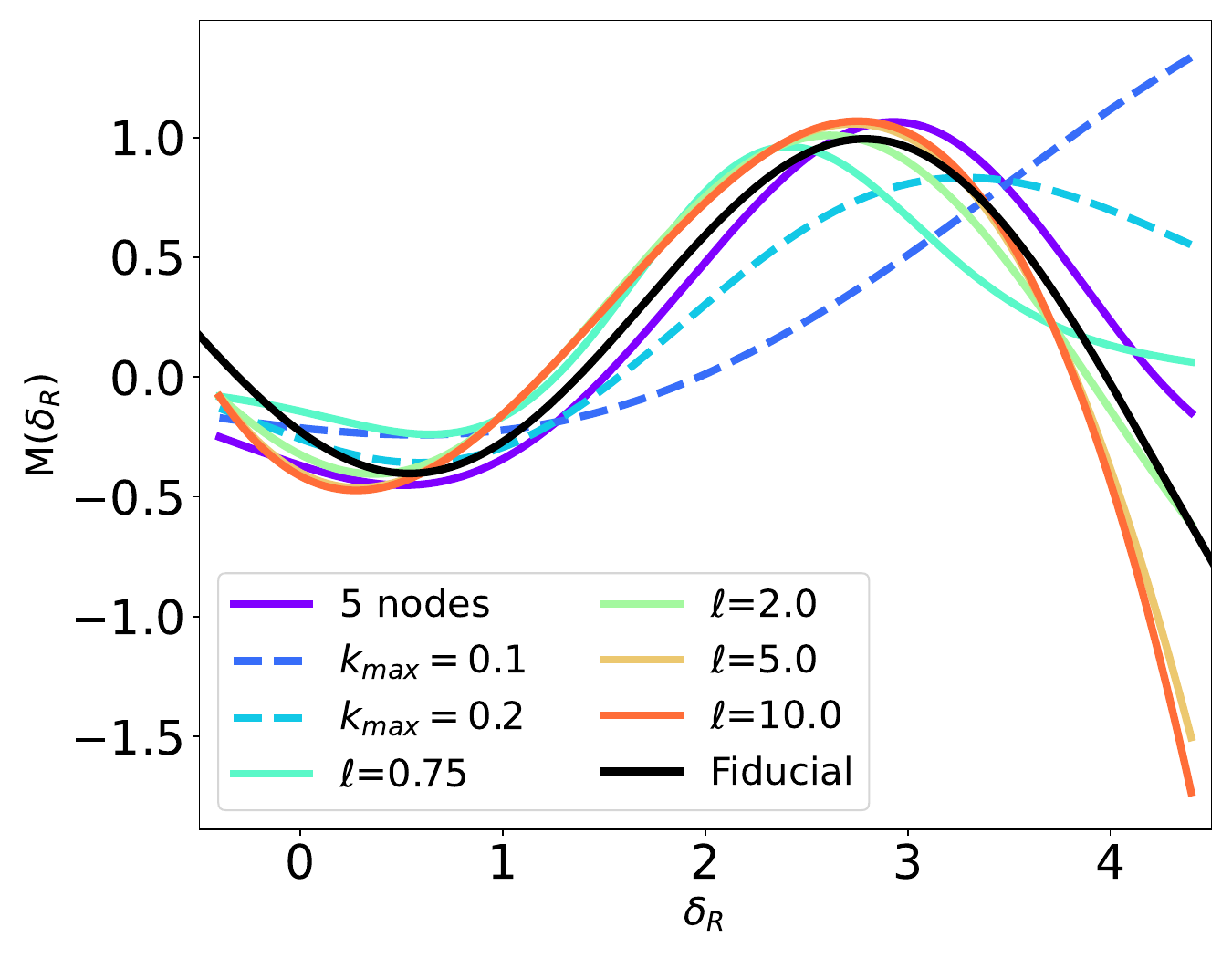}
      \caption{Example of the universality of the mark function. Optimised curves from different tests are displayed, including varying the number of nodes, the length scale of the GPR ($\ell$), and the maximum k used in the analysis.}
      \label{fig:unversality}
    \end{figure}
    As described above, under our fiducial setup we have been able to find an optimal mark function that is qualitatively universal (i.e. functions with similar shapes are able to maximise the information recovered on several cosmological parameters). It is however important to verify that the resulting optimal function is not strongly dependent on the specific choices made in our analysis, particularly regarding the methodology used to describe general functions. To this end, we repeated our analysis changing some of these choices, and compared the resulting optimal mark functions with our fiducial result.

    Specifically, we explored the following choices:
    \begin{itemize}
      \item The {\bf number of nodes} used to describe the Gaussian process. Our fiducial choice of 4 nodes (corresponding to 3 degrees of freedom, after normalising the function as described in Section \ref{sssec:theory.marks.gen} may not allow for sufficient freedom to discover potentially more optimal functions. We thus repeat our analysis for 5 Gaussian process nodes.
      \item The {\bf correlation length} of the Gaussian process $\ell$. This parameter determines the ``smoothness'' of the function (i.e. the speed with which it can vary across its domain of definition). Our fiducial choice of $\ell$ (different for each smoothing scale $R$) was selected as a compromise between the range of variation of $\delta_R$ and the separation between nodes $\Delta\delta_R$\footnote{For instance, small lengths $\ell\ll\Delta\delta_R$ would force the recovered Gaussian process mean to systematically drift towards zero between nodes.}. This choice may have a significant impact on the shape of the recovered optimal mark, restricting its freedom to vary over overdensity ranges $\ll\ell$. We repeat our study for 4 different correlation lengths: $\ell\in\{0.75,\,2.0,\,5.0,\,10\}$.
      \item The {\bf smallest scale} used in the analysis. In our fiducial analysis, we use all Fourier modes up to $k_{\rm max}=0.3\,\impch$. This scale was chosen as a compromise between ensuring that we sufficiently probe the mildly non-linear regime where the overdensity field is clearly non-Gaussian, while avoiding complex astrophysical effects (e.g. baryonic effects) that may dominate the theoretical error budget on small scales ($k\lesssim1\,\impch$). The shape of the optimal mark function likely depends on the range of scales over which non-Gaussian information can be extracted, and therefore we repeat our analysis for $k_{\rm max}=0.2\,\impch$ and $0.1\,\impch$. 
    \end{itemize}

    The results are shown in Fig. \ref{fig:unversality} for the $R=10\,\mpch$ smoothing scale, with our fiducial mark function shown in black. We find that increasing the number of GP nodes (purple line, with node positions marked as circles) leads to an optimal mark that is qualitatively equivalent to our fiducial one. Likewise, varying the GP correlation length does not lead to significant deviations (again, qualitatively), from this optimal mark. We thus conclude that the constraints imposed on the mark function by our analysis, which restricts its degree of smoothness and allowed range of variation, do not have a significant effect on the shape of the optimal mark function found here.
             
    On the other hand, we observe that reducing the maximum wavenumber used in the analysis does have a gradual effect on the shape of the optimal mark seemingly shifting the high-density maximum of the function towards larger values of $\delta_R$. Specifically, we find cutting  $k_{\rm max}=0.2\,\impch$ and $k_{\rm max}=0.1\,\impch$ yields the error improvements in  $\sigma_8$ and $\Omega_m$ of (1.2, 2.4), and  (1.1, 1.1), respectively. These are reduced significantly compared to the fiducial results of (2.1, 4.6). As justified above, this is not entirely surprising, since the threshold scale $k_{\rm max}$ directly constrains the level of non-Gaussianity present in the field on the scales used. This result, however, implies that the exact form of the optimal mark function does depend on some of the analysis choices, and therefore should be re-calibrated whenever these choices change.

\section{Discussion}\label{sec:conc}
  In this work, we studied the use of marked power spectra as a way to enhance the constraining power of large-scale structure analysis for cosmology. In particular, we studied the problem of finding an optimal form for the mark function, beyond that allowed by parametric approaches used in the past. To do so, we have modelled the mark function as the posterior mean of a Gaussian process defined by its value at a number of nodes (4 or 5 in our case) spread throughout the domain of the smoothed density field $\delta_R$. To facilitate this study, we have identified the symmetries of this mark function (i.e. transformations under which the marked field retains the same amount of information), in particular showing that mark functions are equivalent under affine transformations. Besides reducing the number of degrees of freedom to explore in our optimisation, this is also useful in interpreting the properties of different mark functions (e.g. the type of environments over which the original field is up- or down-weighted).

  We find that it is possible to find a single mark function that is simultaneously optimal in the recovery of different cosmological parameters ($\Omega_m$ and $\sigma_8$ in the case studied here), both individually and jointly. The form of this optimal mark function is robust against the main choices made to define it (e.g. the number of GP nodes and the correlation length of the GP covariance). When applied to the matter overdensity field, this mark is able to improve the uncertainties on $\Omega_m$ and $\sigma_8$ by a factor of $\sim 4$ and $\sim 2$, respectively (see Table \ref{tab:optimisers}). The optimal mark functions found for three different smoothing scales are shown in Fig. \ref{fig:optimisers}. Their main common feature seems to be the existence of a minimum around $\delta_R\sim0$ which, making use of the affine symmetry mentioned above, can be interpreted as the function upweighting both over- and under-densities with the same sign (and not opposite signs, as a zero-crossing mark function would imply). This interpretation is only qualitative and we have seen that the exact shape of the function depends on the choices made when defining the characteristic scales of the analysis: namely, the smoothing scale $R$ used to define the environment and the smallest scale used to constrain cosmology, $k_{\rm max}$.

  The results presented here are subject to a number of caveats that must be noted. Perhaps most importantly, our analysis has focused on the use of marked power spectra for the study of the three-dimensional matter overdensity field itself. Real observations instead must rely on biased tracers of this matter overdensity, such as galaxies, or use projected (i.e. 2D) probes, such as cosmic shear or CMB lensing. Our analysis may be extended to 2D fields in the latter case (Cowell+ in prep.), but the use of biased tracers implies significant departures from the assumptions of our study. First, our ability to recover information on cosmological parameters is significantly hampered by the need to marginalise over galaxy bias \citep[or, more in general, the details of galaxy-halo relation,][]{Wechsler_2018}. Although the use of marked fields may improve our ability to break the degeneracy between bias and cosmological parameters, it is not immediately clear that a single mark function may be found that simultaneously recovers optimal constraints on all parameters. Secondly, galaxy clustering studies inevitably suffer from shot noise, caused by the discrete nature of the tracer. This degrades the precision with which information can be extracted from the smallest scales and may therefore affect the performance and the specific shape of the optimal mark function (just as we have shown a change in $k_{\rm max}$ does). Finally, we have not considered the effects of super sample covariance on the shape of the optimal mark~\citep{Bayer2023_SSC}.

  In the same vein, our analysis has been relatively limited in scope, studying only constraints on two cosmological parameters. More than one mark function may be needed to optimise constraints on a larger parameter set, particularly if parameters that rely heavily on the small-scale dependence of the power spectrum, such as neutrino masses, are included. The impact of other parameters that are poorly constrained by large-scale structure, but which show degeneracies with parameters of interest (e.g $h$, or $n_s$), is also unclear. The design of optimal general marks (or sets thereof) for the extraction of constraints from multi-dimensional parameter spaces is an interesting problem that we leave for future work.

  Finally, we have relied on a Gaussian covariance matrix approximation to derive optimal mark functions. This was necessary to avoid having to carry out expensive calculations over hundreds of simulations at each step in the optimisation. As we show in Fig. \ref{fig:cov_plots}, while the Gaussian covariance is a relatively good approximation on large scales, the power spectra involving the marked field display significant departures from it, particularly on small scales. We have argued that these non-Gaussian contributions, which become increasingly relevant for larger smoothing scales, are relatively harmless since they effectively correspond to a larger variance for a single data mode. Nevertheless, the exact shape of the optimal mark is likely sensitive, at some level, to this approximation. The impact of non-Gaussian covariances has also affected our analysis by forcing us to choose a relatively conservative scale cut ($k_{\rm max}=0.3\,\impch$), beyond which other non-Gaussian contributions would become more relevant. As we have seen, this choice can significantly affect the shape of the resulting optimal mark. A deeper study of the covariance matrix for marked fields would be useful, not only to improve the design of optimal mark functions but, perhaps more importantly, to facilitate their use in cosmological analyses without having to rely upon an expensive simulation-based approach to compute them.

  As confirmed by previous works, marked power spectra are a promising avenue to extract non-Gaussian information from large-scale structure probes. Our study has shown that the statistical power of this technique can be significantly enhanced by searching for an optimal mark function, allowing it to take both positive and negative values, and to depart from simple parametric forms. Applying a similar procedure to the various observational LSS probes under exploitation will therefore increase the return on investment in current and next-generation investment. 

\section*{Acknowledgements}
  We would like to thank Francisco Villaescusa-Navarro, Elena Massara and Joaquin Armijo for their useful discussions. JAC is funded by a Kavli/IPMU PhD Studentship. DA acknowledges support from the Beecroft Trust. This work was supported by JSPS KAKENHI Grants 23K13095 and  23H00107 (to JL). We made extensive use of computational resources at the University of Oxford Department of Physics, funded by the John Fell Oxford University Press Research Fund. 

\section*{Data Availability}
  The data and software used for this analysis can be made available upon request.

\bibliographystyle{mnras}
\bibliography{bib} 

\bsp	
\label{lastpage}
\end{document}